\documentclass[aps,prd,nofootinbib,amsmath,amssymb,superscriptaddress,twocolumn,10pt]{revtex4}%superscriptaddress,showpacs,

\pdfoutput=1

\usepackage{graphicx}
\usepackage{dcolumn}
\usepackage{bm}
\usepackage{amssymb}
\usepackage{latexsym}
\usepackage{booktabs}
\usepackage{amsmath}
\usepackage{multirow}
\usepackage[colorlinks=true, linkcolor=red, citecolor=blue]{hyperref}

\newcommand{\be}{\begin{equation}}
\newcommand{\ee}{\end{equation}}
\newcommand{\bq}{\begin{eqnarray}}
\newcommand{\eq}{\end{eqnarray}}

\begin{document}

\title{Constraints on inflation revisited: An analysis including the latest local measurement of the Hubble constant}

\author{Rui-Yun Guo}
\affiliation{Department of Physics, College of Sciences,
Northeastern University, Shenyang 110004, China}
\author{Xin Zhang\footnote{Corresponding author}}
\email{zhangxin@mail.neu.edu.cn}
\affiliation{Department of Physics, College of Sciences,
Northeastern University, Shenyang 110004, China}
\affiliation{Center for High Energy Physics, Peking University, Beijing 100080, China}

\begin{abstract}

We revisit the constraints on inflation models by using the current cosmological observations involving the latest local measurement of the Hubble constant ($H_{0} = 73.00\pm 1.75$ km s $^{-1}$ Mpc$^{-1}$). We constrain the primordial power spectra of both scalar and tensor perturbations with the observational data including the Planck 2015 CMB full data, the BICEP2 and Keck Array CMB B-mode data, the BAO data, and the direct measurement of $H_0$. In order to relieve the tension between the local determination of the Hubble constant and the other astrophysical observations, we consider the additional parameter $N_{\rm eff}$ in the cosmological model. We find that, for the $\Lambda$CDM+$r$+$N_{\rm eff}$ model, the scale invariance is only excluded at the 3.3$\sigma$ level, and $\Delta N_{\rm eff}>0$ is favored at the 1.6$\sigma$ level. Comparing the obtained 1$\sigma$ and 2$\sigma$ contours of $(n_s,r)$ with the theoretical predictions of selected inflation models, we find that both the convex and concave potentials are favored at 2$\sigma$ level, the natural inflation model is excluded at more than 2$\sigma$ level, the Starobinsky $R^2$ inflation model is only favored at around 2$\sigma$ level, and the spontaneously broken SUSY inflation model is now the most favored model.

\end{abstract}

\maketitle
\section{Introduction}

Inflation is the leading paradigm to explain the origin of the primordial density perturbations and the primordial gravitational waves, which is a period of accelerated expansion of the early universe. It can resolve a number of puzzles of the standard cosmology, such as the horizon, flatness, and monopole problems~\cite{Starobinsky1980,Guth: 1981,Linde: 1982,Albrecht: 1982}, and offer the initial conditions for the standard cosmology. During the epoch, inflation can generate the primordial density perturbations, which seeded the cosmic microwave background (CMB) anisotropies  and the large-scale structure (LSS) formation in our universe. Thus, current cosmological observations can be used to explore the nature of inflation. For example, the measurements of CMB anisotropies have confirmed that inflation can provide a nearly  scale-invariant primordial power spectrum~\cite{Hinshaw:2012aka,Sievers:2013ica,Hou:2012xq,Ade:2013ydc}.

Although inflation took place at energy scale as high as $10^{16}$ GeV, where particle physics remains elusive, hundreds of different theoretical scenarios have been proposed. Thus selecting an actual version of inflation has become a major issue in the current study. As mentioned above, the primordial perturbations can lead to the CMB anisotropies and LSS formation, so comparing the predictions of these inflation models with cosmological data can provide the possibility to identify the suitable inflation models.

The astronomical observations measuring the CMB anisotropies have provided an excellent opportunity to explore the physics in the early universe. The Planck collaboration~\cite{Ade:2015xua} has measured the primordial power spectrum of density perturbations with an unprecedented accuracy. Namely, the spectral index is measured to be $n_{s}=0.968\pm 0.006$ ($1\sigma$), ruling out the scale invariance at more than $5\sigma$, and the running of the spectral index is measured to be $d n_{\rm s}/d \ln k =-0.003\pm 0.007$ ($1\sigma$), from the Planck temperature data combined with the Planck lensing likelihood. The constraint on the tensor-to-scalar ratio is $r_{0.002}<0.11$ at the $2\sigma$ level, also derived by using the Planck temperature data combined with the Planck lensing likelihood. In addition, the Keck Array and BICEP2 collaborations~\cite{Ade2016} released a highly significant detection of B-mode polarization with inclusion of the first Keck Array B-mode polarization at 95 GHz. These data were taken by the BICEP2 and Keck Array CMB polarization experiments up to and including the 2014 observing season to improve the current constraints on primordial power spectra. The constraint on the tensor-to-scalar ratio is $r_{0.05}<0.09$ at the $2\sigma$ level from the B-mode only data of BICEP2 and Keck Array. The tighter constraint is $r_{0.05}<0.07$ at the $2\sigma$ level when the BICEP2/Keck Array B-mode data are combined with the Planck CMB data plus other astrophysical observations.

The baryon acoustic oscillation (BAO) data can effectively break the degeneracies between cosmological parameters and further improve the constraints on inflation models (see, e.g., Refs.~\cite{Huang:2015cke,Tram:2016rcw,Gerbino:2016sgw,DiValentino:2016ziq,DiValentino:2016nni}). In this paper, we employ the latest BAO measurements including the Date Release 12 of the SDSS-III Baryon Oscillation Spectroscopic Survey (BOSS DR12)~\cite{Gil-Marin:2015nqa}, the 6dF Galaxy Survey (6dFGS) measurement~\cite{Beutler:2011hx}, and the Main Galaxy Sample of Data Release 7 of Sloan Digital Sky Survey (SDSS-MGS)~\cite{Ross:2014qpa}.

Recently, Riess et al.~\cite{Riess:2016jrr} reported their new result of direct measurement of the Hubble constant, $H_{0} = 73.00\pm 1.75$ km s $^{-1}$ Mpc$^{-1}$, which is $3.3\sigma$ higher than the fitting result, $H_{0} = 66.93\pm 0.62$ km s $^{-1}$ Mpc$^{-1}$, derived by the Planck collaboration~\cite{Aghanim:2016yuo} based on the $\Lambda$CDM model assuming $ \sum m_{\nu}= 0.06$ eV using the Planck TT, TE, EE+lowP data. The strong tension between the new measurement of $H_{0}$ and the Planck data may be from some systematic uncertainties in the measurements or some new physics effects. In order to reconcile the new measurement of $H_{0}$ and the Planck data, one can consider the new physics by adding some extra parameters, such as the parameters describing a dynamical dark energy~\cite{Li:2013dha,Qing-Guo:2016ykt}, extra relativistic degrees of freedom~\cite{Riess:2016jrr,Zhang:2014dxk,Zhang:2014ifa,DiValentino:2016ucb,Benetti:2017gvm} and light sterile neutrinos~\cite{Zhang:2014dxk,Zhang:2014ifa,Zhang:2014nta,Feng:2017nss,Zhao:2017urm,Feng:2017mfs,Zhao:2017jma}.

Although there are strong tensions between the new measurement of $H_{0}$ and other cosmological observations, the result of $H_{0} = 73.00\pm 1.75$ km s $^{-1}$ Mpc$^{-1}$ can play an important role in current cosmology due to its reduced uncertainty from 3.3\% to 2.4\%. In this paper, we combine the new measurement of $H_{0}$ with the Planck data, the BICEP2/Keck Array data and the BAO data to constrain inflation models. The aim of this work is to investigate whether the local determination $H_{0} = 73.00\pm 1.75$ km s $^{-1}$ Mpc$^{-1}$ will have a remarkable influence on constraining the primordial power spectra of scalar and tensor perturbations. In order to relieve the tension between the local determination of the Hubble constant and other astrophysical observations, we decide to consider dark radiation, parametrized by $\Delta N_{\rm eff}$ (defined by $N_{\rm eff}-3.046$), in the cosmological model in our analysis. The constraint results of $(n_s,r)$ will be compared with the theoretical predictions of some typical inflation models to make a model selection analysis.

The structure of the paper is organized as follows. In Sec.~\ref{sec:2}, we briefly introduce the single-field slow-roll inflationary scenario. In Sec.~\ref{sec:3}, we report the results of the constraints on the primordial power spectra with the combination of the Planck data, the BICEP2/Keck Array data, the BAO data and the latest measurement of $H_{0}$. In Sec.~\ref{sec:4}, we compare the constraint results of $(n_s,r)$ with the theoretical predictions of some typical inflationary models and show the impacts of the latest measurement of $H_0$ on the inflation model selection. Conclusion is given in Sec.~\ref{sec:5}.

\section{Slow-roll inflationary scenario}\label{sec:2}

In this paper, we only consider the simplest inflationary scenario within the slow-roll paradigm, for which the accelerated expansion of early universe is driven by a homogeneous, slowly rolling scalar field $\phi$. According to the energy density of the inflaton $\rho_{\phi}=\dot{\phi}^{2}/{2} + V(\phi)$, the Friedmann equation becomes
\begin{equation}\label{2.1}
  H^{2}=\frac{1}{3M^{2}_{\rm pl}}\left[\frac{1}{2}\dot{\phi}^{2}+V(\phi)\right],
\end{equation}
where $H=\dot{a}/a$ (with $a$ the scale factor of the universe) is the Hubble parameter, $M_{\rm pl} = 1/\sqrt{8\pi G}$ is the reduced Planck mass, $V(\phi)$ is the inflaton potential, and the dot denotes the derivative with respect to the cosmic time $t$.

The equation of motion for the inflaton satisfies
\begin{equation}\label{2.2}
  \ddot{\phi}+3H\dot{\phi}+V^{\prime}(\phi)=0,
\end{equation}
where the prime is the derivative with respect to the inflaton $\phi$. Due to the slow-roll approximation, $\dot{\phi}^{2} \ll 0$ and $\ddot{\phi} \ll 0$, Eqs.~(\ref{2.1}) and~(\ref{2.2}) can be reduced to
\begin{equation}\label{2.3}
  H^{2} \approx \frac{V(\phi)}{3M^{2}_{\rm pl}},
\end{equation}
\begin{equation}\label{2.4}
  3H\dot{\phi} \approx -V^{\prime}(\phi).
\end{equation}
Usually, the inflationary universe can be characterized with the slow-roll parameters, which can be defined as
\begin{equation}\label{2.5}
 \epsilon = \frac{M^{2}_{\rm pl}}{2}\left[\frac{V^{\prime}(\phi)}{V(\phi)}\right]^{2},
\end{equation}
\begin{equation}\label{2.6}
  \eta = M^{2}_{\rm pl}\left[\frac{V^{\prime \prime}(\phi)}{V(\phi)}\right],
\end{equation}
\begin{equation}\label{2.7}
  \xi^{2}=\frac{M_{\rm pl}^{4}V^{\prime}(\phi)V^{\prime\prime\prime}(\phi)}{V^{2}(\phi)},
\end{equation}
and so on. The inflaton slowly rolls down its potential $V(\phi)$ as long as $\epsilon \ll 1$ and $|\eta| \ll 1$.

The tensor-to-scalar ratio, which is defined to be the ratio of the tensor spectrum $P_{\rm t}(k)$ to the scalar spectrum $P_{\rm s}(k)$, can be given by the slow-roll approximation as
\begin{equation}\label{2.9}
  r = \frac{P_{\rm t}(k)}{P_{\rm s}(k)} = 16 \epsilon.
\end{equation}

Similarly, according to the slow-roll approximation, we can obtain the spectral index
\begin{equation}\label{2.13}
  n_{\rm s} = 1- 6\epsilon + 2\eta,
\end{equation}
and the running spectral index
\begin{equation}\label{2.14}
 dn_{\rm s}/d\ln k = 16\epsilon \eta -24\epsilon^{2}-2\xi^{2}.
\end{equation}
By constraining these parameters using cosmological observations, we can effectively distinguish between different inflation models.

\section{Constraints on primordial power spectra}\label{sec:3}

\begin{table*}[ht!]\tiny
\caption{ The fitting results of the cosmological parameters in the $\Lambda$CDM+$r$, $\Lambda$CDM+$r$+$N_{\rm eff}$, $\Lambda$CDM+$r$+$d n_{\rm s}/d \ln k$, and $\Lambda$CDM+$r$+$d n_{\rm s}/d \ln k$+$N_{\rm eff}$ models using the Planck+BK+BAO+$H_{0}$ data. }
\label{table}
\small
\setlength\tabcolsep{2.8pt}
\renewcommand{\arraystretch}{1.2}
\centering
\begin{tabular}{cccccccccccccccccc}
\\
\hline\hline
Parameter & $\Lambda$CDM+$r$  & $\Lambda$CDM+$r$+$N_{\rm eff}$  & $\Lambda$CDM+$r$+$d n_{\rm s}/d \ln k$  &   $\Lambda$CDM+$r$+$d n_{\rm s}/d \ln k$+$N_{\rm eff}$  \\
\cline{1-5}

$\Omega_{\rm b}h^2$   &$0 .02238\pm0.00014$
                      &$0.02253\pm0.00017$
                      &$0 .02242\pm0.00015$
                      &$0 .02253\pm0.00019$\\

$\Omega_{\rm c}h^2$   &$0 .1177\pm0.0010$
                      &$0.1216\pm0.0027$
                      &$0 .1174\pm0.0010$
                      &$0 .1218\pm0.0030$\\

$100\theta_{\rm MC}$ &$1 .04106\pm0.00029$
                     &$1.04062^{+0.00039}_{-0.00038}$
                     &$1 .04109^{+0.00029}_{-0.00030}$
                     &$1 .04061\pm0.00042$\\

$\tau$               &$0 .075\pm0.012$
                     &$0.074\pm0.012$
                     &$0 .077\pm0.012$
                     &$0 .074\pm0.014$\\

${\rm{ln}}(10^{10}A_{\rm s})$   &$3 .079\pm0.023$
                            &$3.088\pm0.023$
                            &$3 .084\pm0.023$
                            &$3 .087\pm0.027$\\

$n_{\rm s}$                &$0 .9699^{+0.0040}_{-0.0039}$
                           &$0.9787^{+0.0064}_{-0.0065}$
                           &$0 .9701^{+0.0041}_{-0.0042}$
                           &$0 .9781\pm0.0080$\\

\hline
$d n_{\rm s}/d \ln k$      &...
                           &...
                           &$-0.0042^{+0.0067}_{-0.0066}$
                           &$0 .0010^{+0.0074}_{-0.0073}$\\

$r_{\rm 0.002}$ ($2\sigma$) &$<0.069$
                            &$<0.071$
                            &$<0.077$
                            &$<0.074$\\

$N_{\rm eff}$              &...
                           &$3.30\pm0.16$
                            &...
                           &$3 .30\pm0.18$\\

\hline

$\Omega_{\rm m}$           &$0 .3023\pm0.0060$
                           &$0.2988\pm0.0060$
                           &$0 .3007^{+0.0060}_{-0.0061}$
                           &$0 .2988^{+0.0085}_{-0.0094}$\\

$H_0$                      &$68 .23^{+0.47}_{-0.46}$
                           &$69.63\pm0.99$
                           &$68 .37^{+0.47}_{-0.50}$
                           &$69 .70^{+1.20}_{-1.30}$\\

$\sigma_8$                 &$0 .8188^{+0.0087}_{-0.0085}$
                           &$0.8300\pm0.0110$
                           &$0 .8193^{+0.0085}_{-0.0086}$
                           &$0 .8310\pm0.0120$\\

\hline
$\chi^{2}_{\rm min}$    & $13616.988$
                        & $13612.184$
                        & $13615.324$
                        & $13611.122$
\\
\hline\hline
\end{tabular}
\end{table*}

In this section, we make a comprehensive analysis of constraining the primordial power spectra of scalar and tensor perturbations by combining the new measurement of the Hubble constant, $H_{0} = 73.00\pm 1.75$ km s $^{-1}$ Mpc$^{-1}$~\cite{Riess:2016jrr}, with the Planck data, the BICEP2/Keck Array data and the BAO data, to investigate how the new measurement of $H_{0}$ affects the constraint results of inflation models. We employ the Planck CMB 2015 data set including the temperature power spectrum (TT), the polarization power spectrum (EE), the cross-correlation power spectrum of temperature and polarization (TE), and the Planck low-$\ell$ ($\ell \leq 30$) likelihood (lowP), as well as the lensing reconstruction, which is abbreviated as ``Planck". We employ all the BICEP2 and Keck Array B-mode data with inclusion of 95 GHz band, abbreviated as ``BK". The BAO data include the CMASS and LOWZ samples from the BOSS DR12 at $z_{\rm eff} = 0.57$ and $z_{\rm eff} = 0.32$ \cite{Gil-Marin:2015nqa}, the 6dFGS measurement at $z_{\rm eff} = 0.106$ \cite{Beutler:2011hx}, and the SDSS-MGS measurement at $z_{\rm eff} = 0.15$~\cite{Ross:2014qpa}, abbreviated as ``BAO".

The primordial power spectra of scalar and tensor perturbations can be expressed as
\begin{equation}\label{3.1}
  P_{\rm s}(k) = A_{\rm s}\left(\frac{k}{k_{*}}\right)^{n_{\rm s}-1 +\frac{1}{2} \frac{dn_{\rm s}}{d\ln k} \ln \left(\frac{k}{k_{*}}\right)},
\end{equation}
\begin{equation}\label{3.2}
   P_{\rm t}(k) = A_{\rm t}\left(\frac{k}{k_{*}}\right)^{n_{\rm t}+\frac{1}{2}\frac{dn_{\rm t}}{d\ln k}\ln \left(\frac{k}{k_{*}}\right)},
\end{equation}
where $A_{\rm s}$ and $A_{\rm t}$ correspond to the scalar and tensor amplitudes at the pivot scale $k_{*}$, respectively. For the canonical single-field slow-roll inflation model without the inclusion of the running of the spectral index, we have the consistency relation $n_{\rm t} = -r/8$. When the running spectral index is considered, we then have $n_{\rm t}=-r(2-r/8-n_{\rm s})/8$ and $dn_{\rm t}/d\ln k=r(r/8+n_{\rm s}-1)/8$. We uniformly set the pivot scale as $k_{*} = 0.002$ Mpc$^{-1}$ in this work.

\begin{figure}[ht!]
\begin{center}
\includegraphics[width=8cm]{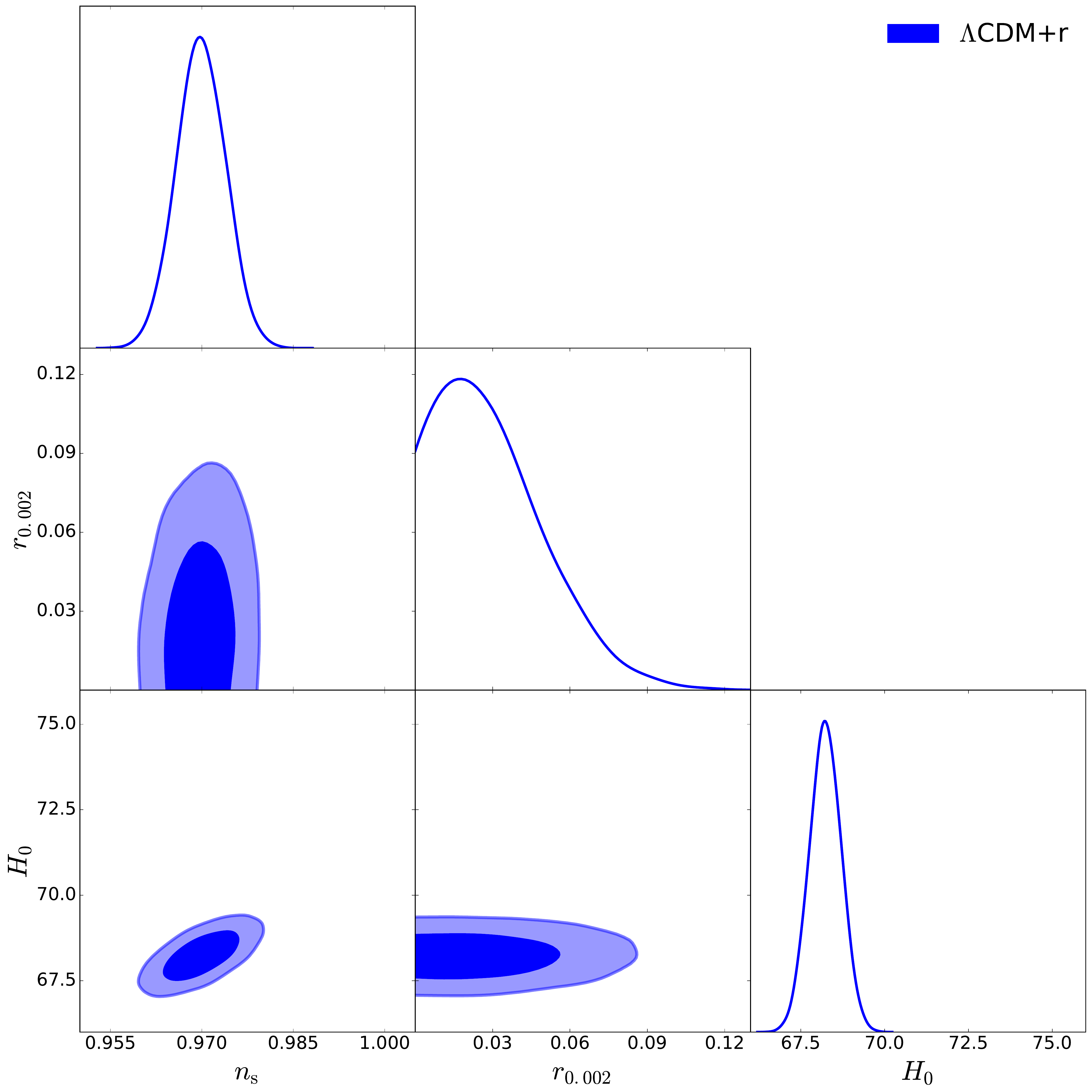}
\end{center}
\caption{One-dimensional marginalized distributions and two-dimensional contours ($1\sigma$ and $2\sigma$) for parameters $n_{\rm s}$, $r_{\rm 0.002}$ and $H_0$ in the $\Lambda$CDM+$r$ model using the Planck+BK+BAO+$H_{0}$ data.}
\label{fig1}
\end{figure}

\begin{figure}[ht!]
\begin{center}
\includegraphics[width=8cm]{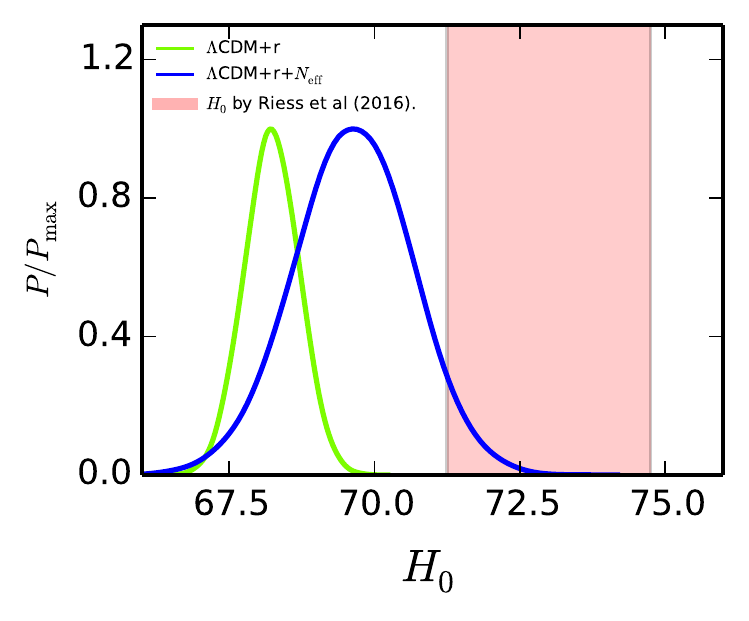}
\end{center}
\caption{The one-dimensional posterior distributions for the parameter $H_{0}$ in the $\Lambda$CDM+$r$ and $\Lambda$CDM+$r$+$N_{\rm eff}$ models using the Planck+BK+BAO+$H_{0}$ data. The light red band denotes the new local measurement of $H_{0}$~\cite{Riess:2016jrr}.}
\label{fig2}
\end{figure}

\begin{figure*}[ht!]
\begin{center}
\includegraphics[width=10cm]{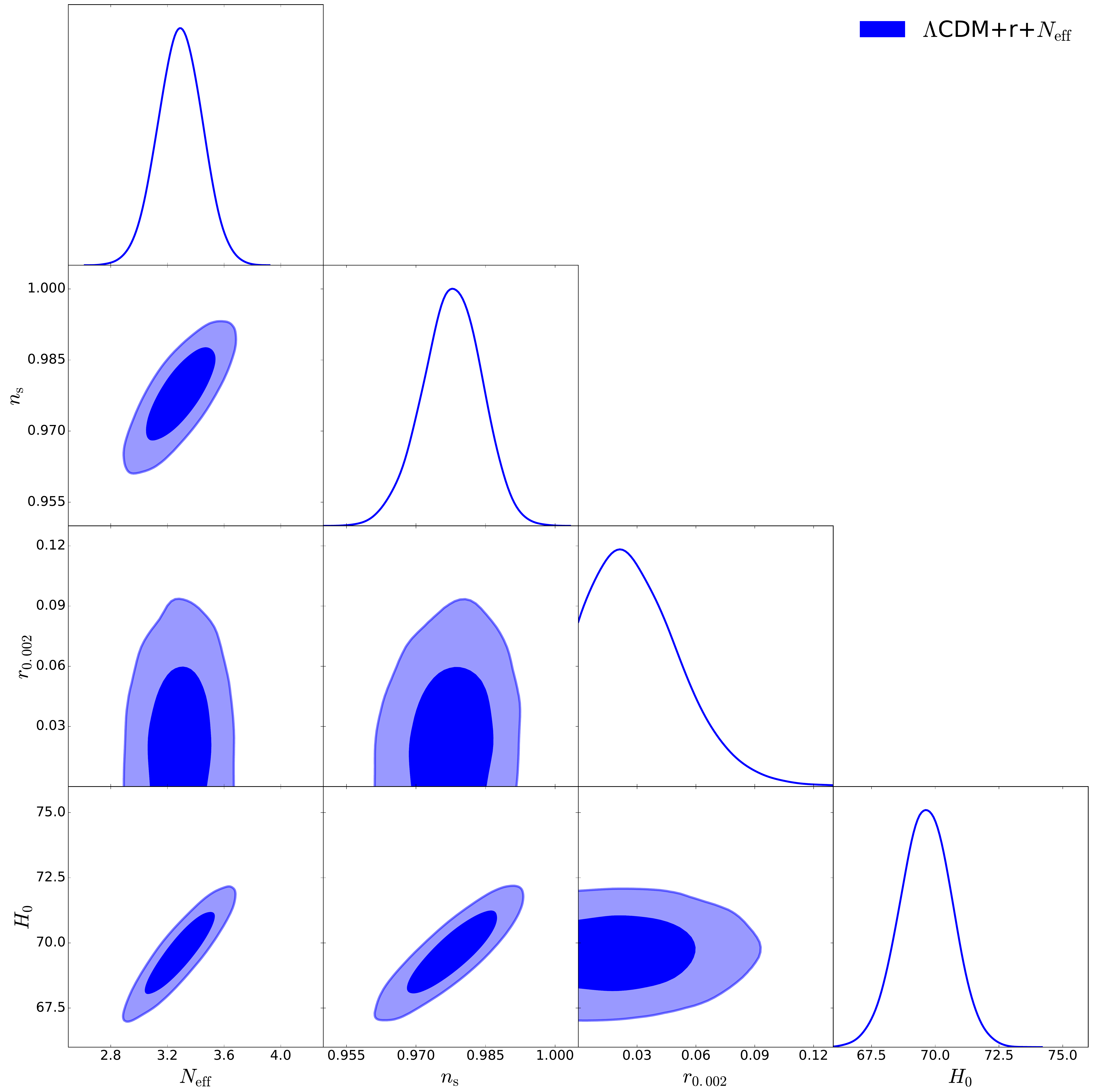}
\end{center}
\caption{One-dimensional marginalized distributions and two-dimensional contours ($1\sigma$ and $2\sigma$) for parameters $N_{\rm eff}$, $n_{\rm s}$, $r_{\rm 0.002}$, and $H_0$ in the $\Lambda$CDM+$r$+$N_{\rm eff}$ model using the Planck+BK+BAO+$H_{0}$ data.}
\label{fig3}
\end{figure*}

\begin{figure*}[ht!]
\begin{center}
\includegraphics[width=11cm]{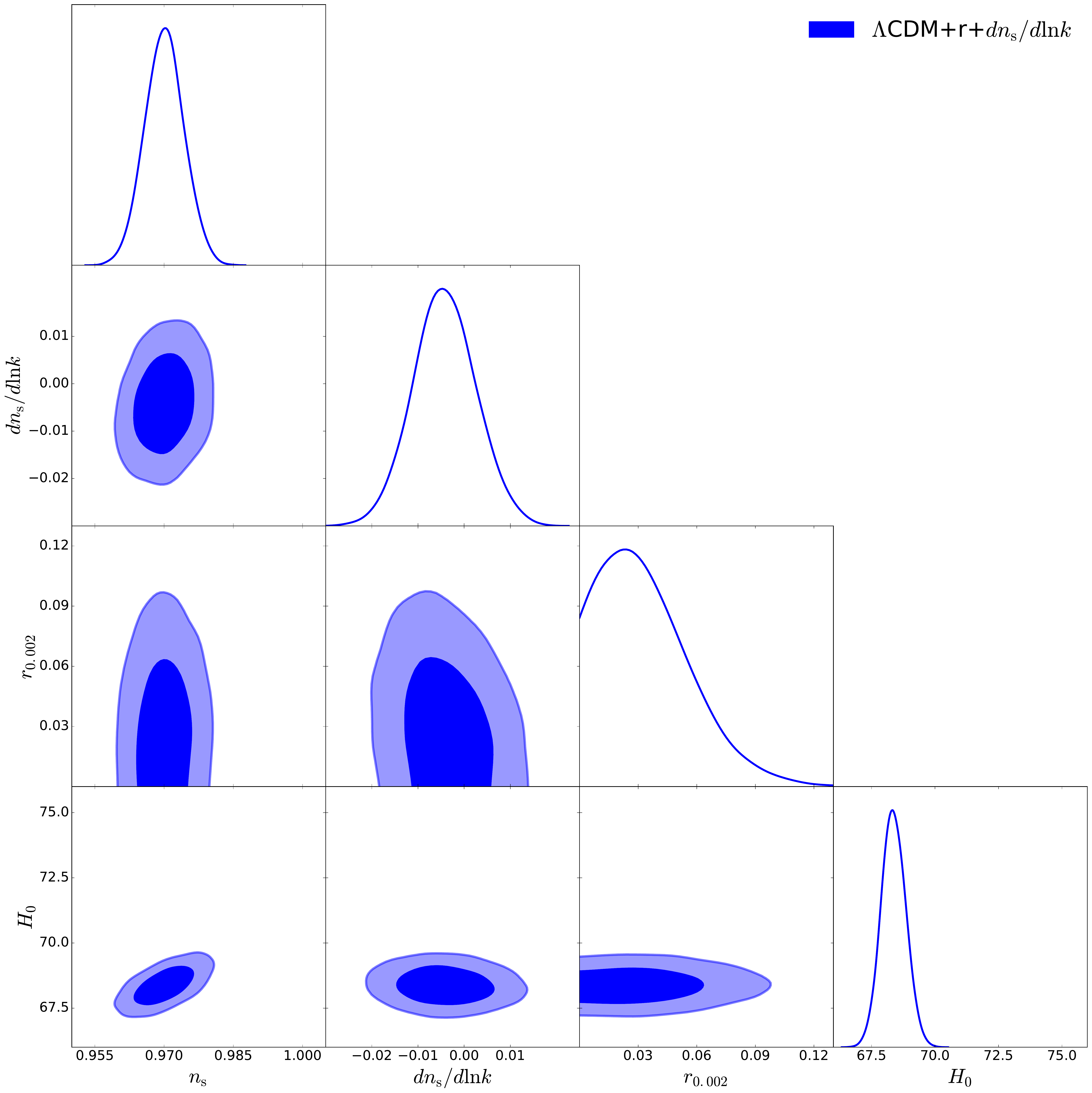}
\end{center}
\caption{One-dimensional marginalized distributions and two-dimensional contours ($1\sigma$ and $2\sigma$) for parameters $n_{\rm s}$, $d n_{\rm s}/d \ln k$, $r_{\rm 0.002}$, and $H_0$ in the $\Lambda$CDM+$r$+$d n_{\rm s}/d \ln k$ model using the Planck+BK+BAO+$H_{0}$ data.}
\label{fig4}
\end{figure*}

\begin{figure*}[ht!]
\begin{center}
\includegraphics[width=11cm]{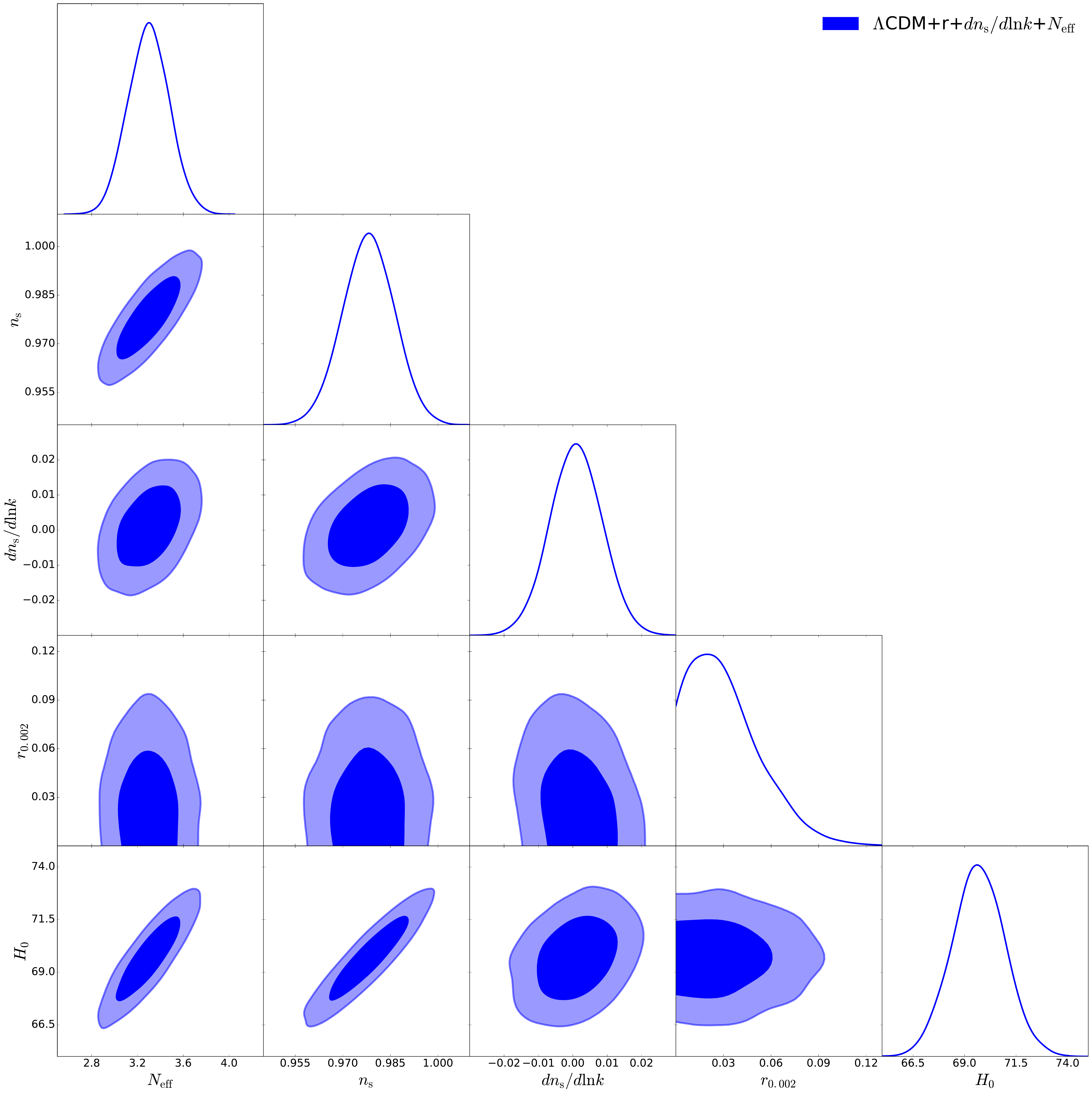}
\end{center}
\caption{One-dimensional marginalized distributions and two-dimensional contours ($1\sigma$ and $2\sigma$) for parameters $N_{\rm eff}$, $n_{\rm s}$, $d n_{\rm s}/d \ln k$, $r_{\rm 0.002}$, and $H_0$ in the $\Lambda$CDM+$r$+$d n_{\rm s}/d \ln k$+$N_{\rm eff}$ model using the Planck+BK+BAO+$H_{0}$ data.}
\label{fig5}
\end{figure*}

There are seven independent free parameters in the base $\Lambda$CDM+$r$ model:
$${\bf P}=\{\Omega_{\rm b} h^2, \Omega_{\rm c} h^2, 100\theta_{\rm MC}, \tau, \ln (10^{10}A_{\rm s}), n_{\rm s}, r\},$$
where $\Omega_{\rm b}h^{2}$ and $\Omega_{\rm c}h^{2}$ denote the present-day densities of baryon and cold dark matter; $\theta_{\rm MC}$ denotes the ratio of the sound horizon $r_{\rm s}$ to the angular diameter distance $D_{\rm A}$ at the last-scattering epoch; $\tau$ denotes the optical depth to reionization; $A_{\rm s}$ and $n_{\rm s}$ denote the amplitude and the spectral index of the primordial power spectra of scalar perturbations, respectively; $r$ denotes the the tensor-to-scalar ratio. When the running is considered, the parameter $d n_{\rm s}/d \ln k$ is added to the cosmological model. In this work, we derive the posterior parameter probabilities by using the Markov Chain Monte Carlo (MCMC) sampler {\tt CosmoMC}~\cite{Lewis:2002ah}.

In Fig.~\ref{fig1}, we give one-dimensional marginalized distributions and two-dimensional contours ($1\sigma$ and $2\sigma$) for the parameters $n_{\rm s}$, $r_{\rm 0.002}$ and $H_0$ in the $\Lambda$CDM+$r$ model using the Planck+BK+BAO+$H_{0}$ data. The constraint results of the $\Lambda$CDM+$r$ model are summarized in the second column of Table~\ref{table}. Here we quote $\pm 1\sigma$ limits for every parameter in the $\Lambda$CDM+$r$ model, except for $r$, which is quoted with the $2\sigma$ upper limit. We obtain the constraints on $r$ and $n_{\rm s}$:
$$
\small
\left.
\begin{array}{c}
r_{0.002}<0.069~(2\sigma) \\
\\
n_{\rm s} = 0 .9699^{+0.0040}_{-0.0039}~(1\sigma)
\end{array}
\right\}\quad\mbox{$\Lambda$CDM+$r$}.
$$
The result of $n_{\rm s}$ for the primordial power spectrum of scalar perturbations excludes the Harrison-Zel'dovich (HZ) scale-invariant spectrum with $n_{\rm s} = 1$ at the $7.5\sigma$ level.

In addition, the constraint on the Hubble constant is $H_{0} = 68 .23^{+0.47}_{-0.46}$ km s$^{-1}$ Mpc$^{-1}$, which is $2.6\sigma$ less than the local determination $H_{0} = 73.00\pm 1.75$ km s$^{-1}$ Mpc$^{-1}$. Namely, the direct measurement of $H_{0} = 73.00\pm 1.75$ km s$^{-1}$ Mpc$^{-1}$ is in tension with the fit result derived by the Planck+BK+BAO+$H_{0}$ data based on the $\Lambda$CDM+$r$ model. As shown in Fig.~\ref{fig2}, the green line denotes the one-dimensional posterior distribution for the parameter $H_{0}$ in the $\Lambda$CDM+$r$ model using the Planck+BK+BAO+$H_{0}$ data, and the light red band denotes the new local measurement of $H_{0}$. Obviously, there is a strong tension between the two results.

Next, we consider the extra relativistic degrees of freedom (i.e., the additional parameter $N_{\rm eff}$) in the cosmological model to relieve the tension between the latest measurement of $H_{0}$ and other observational data. The total radiation energy density in the universe is given by
\begin{equation}\label{3.3}
  \rho_r = \left[1+N_{\rm eff}\frac{7}{8}\left(\frac{4}{11}\right)^{4/3}\right] \rho_{\gamma},
\end{equation}
where $\rho_{\gamma}$ is the energy density of photons. If there are only three-species active neutrinos in the universe, we have the standard value of $N_{\rm eff} = 3.046$. Any additional value of $\Delta N_{\rm eff} = N_{\rm eff} - 3.046 > 0$ indicates the existence of some dark radiation in the universe. Now, we follow Planck collaboration \cite{Ade:2015xua} to constrain $N_{\rm eff}$ as a free parameter, varying within its prior range of $[0, 6]$. Values of $N_{\rm eff}<3.046$ are less well motivated, because such values would require that standard neutrinos are incompletely thermalized or additional photons are produced after the neutrino decoupling, but we still include this range for completeness.

The third column of Table~\ref{table} gives the constraint results of the cosmological parameters in the $\Lambda$CDM+$r$+$N_{\rm eff}$ model using the Planck+BK+BAO+$H_{0}$ data. We obtain the constraints on $r$ and $n_{\rm s}$:
$$
\small
\left.
\begin{array}{c}
r_{0.002} < 0.071~(2\sigma) \\
\\
n_{\rm s}=0.9787^{+0.0064}_{-0.0065}~(1\sigma)
\end{array}
\right\} \quad\mbox{$\Lambda$CDM+$r$+$N_{\rm eff}$}.
$$
The value of $n_{\rm s}$ becomes larger than that without considering $N_{\rm eff}$. The fit result of $N_{\rm eff} = 3.30\pm0.16$ indicates that $\Delta N_{\rm eff} > 0$ is favored at the $1.6\sigma$ level. Due to a positive correlation between $n_{\rm s}$ and $N_{\rm eff}$, as shown in Fig.~\ref{fig3}, $\Delta N_{\rm eff} > 0$ will lead to a larger $n_{\rm s}$.

On the other hand, a larger Hubble constant, $H_{0} = 69.63\pm0.99$ km s$^{-1}$ Mpc$^{-1}$, is obtained when the parameter $N_{\rm eff}$ is considered, which is only $1.7\sigma$ less than the local determination $H_{0} = 73.00\pm 1.75$ km s$^{-1}$ Mpc$^{-1}$. Namely, the tension between $H_{0} = 73.00\pm 1.75$ km s$^{-1}$ Mpc$^{-1}$ and other observational data is greatly alleviated by introducing the parameter $N_{\rm eff}$ in the cosmological model. As showed in Fig.~\ref{fig2}, the constraint on $H_{0}$ derived using the Planck+BK+BAO+$H_{0}$ data in the $\Lambda$CDM+$r$+$N_{\rm eff}$ model is much closer to the local measurement of $H_{0}$. In addition, when the free parameter $N_{\rm eff}$ is included in the cosmological model, $\chi^{2}$ decreases from $13616.988$ to $13612.184$. The big $\chi^2$ difference, $\Delta \chi^{2}=-4.804$, implies that the $\Lambda$CDM+$r$+$N_{\rm eff}$ model, compared to the $\Lambda$CDM+$r$ model, is more favored by the current Planck+BK+BAO+$H_{0}$ data. Here we note that in this paper we compare models through only a $\chi^2_{\rm min}$ comparison, because we constrain these models using the same data combination. In this situation, if one additional parameter can lead to $\chi^2_{\rm min}$ decreasing by more than 2, then we say that adding this parameter is reasonable statistically. Thus, we do not employ Bayesian information criterion or Bayesian evidence in this paper, since a $\chi^2_{\rm min}$ comparison is sufficient for our task.

Furthermore, we consider the inclusion of the running of the spectral index, $d n_{\rm s}/d \ln k$, in the fit to the Planck+BK+BAO+$H_{0}$ data. Figure~\ref{fig4} gives one-dimensional marginalized distributions and two-dimensional contours ($1\sigma$ and $2\sigma$) for parameters $n_{\rm s}$, $d n_{\rm s}/d \ln k$, $r_{\rm 0.002}$, and $H_0$ in the $\Lambda$CDM+$r$+$d n_{\rm s}/d \ln k$ model using the Planck+BK+BAO+$H_{0}$ data. We obtain the constraints on $r$, $n_{\rm s}$ and $d n_{\rm s}/d \ln k$ (see also the fourth column in Table~\ref{table}):
$$
\small
\left.
\begin{array}{c}
r_{0.002} < 0.077~(2\sigma) \\
\\
n_{\rm s}=0 .9701^{+0.0041}_{-0.0042}~(1\sigma) \\
\\
d n_{\rm s}/d \ln k=-0.0042^{+0.0067}_{-0.0066}~(1\sigma)
\end{array}
\right\} \quad\mbox{$\Lambda$CDM+$r$+$d n_{\rm s}/d \ln k$}.
$$
We find that $d n_{\rm s}/d \ln k=0$ is well consistent with the Planck+BK+BAO+$H_{0}$ data in this case, and the fit result $H_{0}=68 .37^{+0.47}_{-0.50}$ km s$^{-1}$ Mpc$^{-1}$ is still in tension with the direct $H_{0}$ measurement. The comparison with the $\Lambda$CDM+$r$ model gives $\Delta\chi^2=-1.664$, implying that adding the parameter $d n_{\rm s}/d \ln k$ does not effectively improve the fit. The comparison with the $\Lambda$CDM+$r$+$N_{\rm eff}$ model gives $\Delta\chi^2=3.14$, explicitly showing that $N_{\rm eff}$ is much more worthy to be added than $d n_{\rm s}/d \ln k$ in the sense of improving the fit.

In Fig.~\ref{fig5}, we give one-dimensional marginalized distributions and two-dimensional contours ($1\sigma$ and $2\sigma$) for the parameters $N_{\rm eff}$, $n_{\rm s}$, $d n_{\rm s}/d \ln k$, $r_{\rm 0.002}$, and $H_0$ in the $\Lambda$CDM+$r$+$d n_{\rm s}/d \ln k$+$N_{\rm eff}$ model using the Planck+BK+BAO+$H_{0}$ data.  We obtain the constraints on $r$, $n_{\rm s}$ and $d n_{\rm s}/d \ln k$ (see also the last column in Table~\ref{table}):
$$
\small
\left.
\begin{array}{c}
r_{0.002} < 0.074~(2\sigma) \\
\\
n_{\rm s}=0 .9781\pm0.0080~(1\sigma) \\
\\
d n_{\rm s}/d \ln k=0 .0010^{+0.0074}_{-0.0073}~(1\sigma)
\end{array}
\right\} \quad\mbox{$\Lambda$CDM+$r$+$d n_{\rm s}/d \ln k$+$N_{\rm eff}$}.
$$
We find that the fitting results are almost unchanged comparing to the case of the $\Lambda$CDM+$r$+$N_{\rm eff}$ model (although the parameter space is slightly amplified), as shown in the third and fifth columns of Table~\ref{table}. The results explicitly show that $d n_{\rm s}/d \ln k=0$ is in good agreement with the current observations. A $\chi^2$ comparison shows that, when the additional parameter $d n_{\rm s}/d \ln k$ is included, the $\chi^2_{\rm min}$ value decreases only by 1.062 (i.e., $\Delta \chi^{2}=-1.062$), which implies that the running of the spectral index $d n_{\rm s}/d \ln k$ is not deserved to be considered in the cosmological model in the sense of statistical significance.

\section{Inflation model selection}\label{sec:4}

\begin{figure}[ht!]
\begin{center}
\includegraphics[width=8cm]{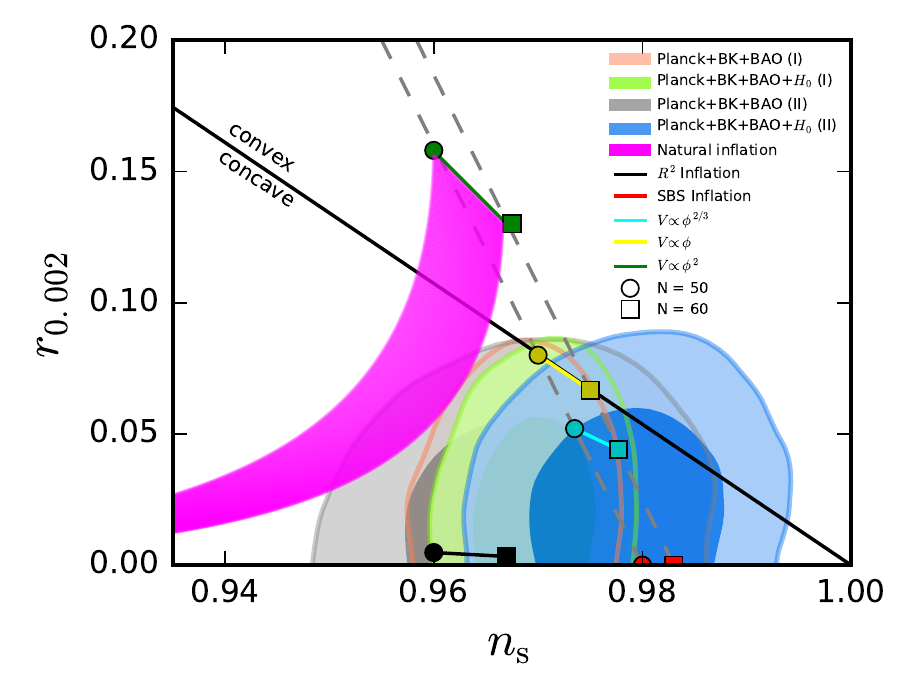}
\end{center}
\caption{Two-dimensional contours ($1\sigma$ and $2\sigma$) for $n_{\rm s}$ and $r_{\rm 0.002}$ using the Planck+BK+BAO and Planck+BK+BAO+$H_{0}$ data, compared to the theoretical predictions of selected inflation models. (I) and (II) correspond to the constraints on the $\Lambda$CDM+$r$ and $\Lambda$CDM+$r$+$N_{\rm eff}$ models, respectively. }
\label{fig6}
\end{figure}

In this section, we consider a few simple and representative inflation models and compare them with the constraint results given in the former section. See also Ref.~\cite{Zhang:2017epd} for a preliminary research. In what follows, we give the predictions of these inflation models for $r$ and $n_{\rm s}$. For these inflation models, we uniformly take the number of $e$-folds $N\in [50, 60]$. In principle, adding the parameter $N_{\rm eff}$ modifies the radiation density and thereby changes the post-inflationary expansion history, so that the $e$-folding number $N$ becomes dependent on the value of $N_{\rm eff}$. However, practically it is hard to link $N$ to the actual observations. Thus, the usual treatment of considering $N\in [50, 60]$ is of course applicable for our analysis.

%Here we have no analysis of the predictions for $d n_{\rm s}/d \ln k$, since the foregoing discusses show $d n_{\rm s}/d \ln k$ is not favored by current cosmological observations. For these inflation models, we uniformly take the number of $e$-folds $N\in [50, 60]$.

The simplest class of inflation models has a monomial potential $V(\phi) \propto \phi^{n}$~\cite{Linde1983}, which is the prototype of the chaotic inflation model. They have the predictions:
\begin{equation}\label{4.1}
  r = \frac{4n}{N},
\end{equation}
\begin{equation}\label{4.2}
  n_{\rm s} = 1- \frac{n+2}{2N},
\end{equation}
where $n$ is any positive number. We take $n= 2/3$, 1, and 2 as typical examples in this work. See also Refs.~\cite{Silverstein:2008sg,McAllister:2008hb,Marchesano:2014mla,McAllister:2014mpa} for relevant studies of this class of models.

The natural inflation model has the effective one-dimensional potential $V(\phi) = \Lambda^{4}(1+\cos (\phi/f))$~\cite{Adams:1992bn,Freese:1993bc}, with the predictions:
\begin{equation}\label{4.3}
  r = \frac{8}{(f/M_{\rm pl})^{2}} \frac{1 + \cos \theta_{N}}{1 - \cos \theta_{N}},
\end{equation}
\begin{equation}\label{4.4}
  n_{\rm s} = 1- \frac{1}{(f/M_{\rm pl})^{2}} \frac{3 + \cos \theta_{N}}{1 - \cos \theta_{N}},
\end{equation}
where $\theta_N$ is given by
\begin{equation}\label{4.5}
 \cos \frac{\theta_{N}}{2} = \exp \left(- \frac{N}{2(f/M_{\rm pl})^{2}}\right).
\end{equation}
Note that different values of $n_{\rm s}$ and $r$ result from the different decay constant $f$ when the number of $e$-folds $N$ is set to be a certain value.

The spontaneously broken SUSY (SBS) inflation model has the potential $V(\phi)= V_{0}(1+c\ln (\phi/Q))$ (where $V_{0}$ is dominant and the parameter $c \ll 1$)~\cite{Copeland:1994vg,Dvali:1994ms,Stewart:1994ts,Binetruy:1996xj,Lyth:1998xn}, with the predictions:
\begin{equation}\label{4.6}
  r \simeq 0,
\end{equation}
\begin{equation}\label{4.7}
  n_{\rm s} = 1 - \frac{1}{N}.
\end{equation}

The Starobinsky $R^{2}$ inflation model is described by the action $S= \frac{M_{\rm pl}^{2}}{2}\int d^{4}x \sqrt{-g} (R + R^{2}/6M^{2})$ (where $M$ denotes an energy scale)~\cite{Starobinsky1980}, with the predictions:
\begin{equation}\label{4.8}
  r \simeq \frac{12}{N^{2}},
\end{equation}
\begin{equation}\label{4.9}
  n_{\rm s} = 1 - \frac{2}{N}.
\end{equation}

In Fig.~\ref{fig6}, we plot two-dimensional contours ($1\sigma$ and $2\sigma$) for $n_{\rm s}$ and $r_{\rm 0.002}$ using the Planck+BK+BAO and Planck+BK+BAO+$H_{0}$ data, compared to the theoretical predictions of selected inflation models. The orange contours denote the constraints on the $\Lambda$CDM+$r$ model with the Planck+BK+BAO data, the green contours denote the constraints on the $\Lambda$CDM+$r$ model with the Planck+BK+BAO+$H_{0}$ data, the gray contours denote the constraints on the $\Lambda$CDM+$r$+$N_{\rm eff}$ model with the Planck+BK+BAO data, and the blue contours denote the constraints on the $\Lambda$CDM+$r$+$N_{\rm eff}$ with the Planck+BK+BAO+$H_{0}$ data.

Comparing the orange and green contours, we find that when the direct measurement of $H_0$ is included in the data combination, the constraint on the $\Lambda$CDM+$r$ model is only changed a little, i.e., a little right shift of $n_s$ is yielded, which does not greatly change the result of inflation model selection (see also Ref.~\cite{Huang:2015cke} for the case of orange contours). According to both the cases of orange and green contours, the inflation model with a convex potential is not favored; both the inflation model with a monomial potential ($\phi$ and $\phi^{2/3}$ cases) and the natural inflation model are marginally favored at around the 2$\sigma$ level; the SBS inflation model is located at out of the 2$\sigma$ region; the Starobinsky $R^2$ inflation model is the most favored model in this case.

When the parameter $N_{\rm eff}$ is considered in the analysis, and if the $H_0$ measurement is not used (i.e., using the Planck+BK+BAO data), we find that the parameter space is greatly amplified (mainly for $n_s$). Comparing the orange and gray contours, we find that without using the $H_0$ measurement the addition of $N_{\rm eff}$ can only amplify the range of $n_s$ but cannot lead to an obvious right shift of $n_s$.

When the $H_0$ measurement is also used, comparing the gray and blue contours, we see that the addition of the $H_0$ prior in the combination of data sets for constraining the $\Lambda$CDM+$r$+$N_{\rm eff}$ model leads to a considerable right shift of $n_s$ (and also a slight shrink of width for the range of $n_s$). In Fig.~\ref{fig3}, we explicitly show that $H_0$ is positively correlated with $N_{\rm eff}$ and $N_{\rm eff}$ is positively correlated with $n_s$, which well explains why the $H_0$ prior (with a larger value of $H_0$) will lead to a larger value of $n_s$ in a cosmological model with $N_{\rm eff}$.

Next, we compare the green and blue contours, which is for the comparison of the $\Lambda$CDM+$r$ and $\Lambda$CDM+$r$+$N_{\rm eff}$ models with the Planck+BK+BAO+$H_{0}$ data, and we see that using the same data sets including the $H_0$ measurement, the consideration of $N_{\rm eff}$ yields a tremendous right shift of $n_s$ (see also Ref.~\cite{Tram:2016rcw}), which largely changes the result of the inflation model selection. As discussed in the last section, the $\Lambda$CDM+$r$+$N_{\rm eff}$ model is much better than the $\Lambda$CDM+$r$ model for the fit to the current Planck+BK+BAO+$H_{0}$ data, since the inclusion of $N_{\rm eff}$ makes the tension between $H_0$ measurement and other observations be greatly relieved and also leads to a much better fit (i.e., the $\chi^2_{\rm min}$ value is largely reduced).

We now compare the predictions of the above typical inflation models with the fit results of $(n_s,r)$ corresponding to the blue contours. We see that, in this case, neither the concave potential nor the convex potential is excluded by the current data. But, it seems that, when comparing the two, the inflation model with the concave potential is more favored by the data. The natural inflation model is now excluded by the data at more than the 2$\sigma$ level. For the inflation models with a monomial potential, we find that the $\phi^2$ model is entirely excluded, the $\phi$ model is only marginally favored (at the edge of the 2$\sigma$ region), and the $\phi^{2/3}$ model is still well consistent with the current data (located in the 1$\sigma$ region). Now, the Starobinsky $R^2$ inflation model is not well favored, because it is located at the edge of the 2$\sigma$ region and actually the $N=50$ point even lies out of the 2$\sigma$ region. We find that in this case the most favored model is the SBS inflation model, which locates near the center of the contours.

Actually, the brane inflation model is also well consistent with the current data in this case (for previous analyses of brane inflation, see, e.g., Refs.~\cite{Ma:2008rf,Ma:2013xma}). We leave a comprehensive analysis for the brane inflation model to a future work.

From the analysis in this paper, we have found that the inclusion of the latest local measurement of the Hubble constant can exert significant influence on the model selection of inflationary models, but one must be aware of that the result is dependent on the assumption of dark radiation in the cosmological model. Without the addition of the parameter $N_{\rm eff}$, the $H_0$ measurement is in tension with the Planck observation, and the $H_0$ prior actually does not greatly influence the fit result of the primordial power spectra (see the comparison of the orange and green contours in Fig.~\ref{fig6}). The $H_0$ tension can be largely relieved provided that the parameter $N_{\rm eff}$ is considered in the model (the tension is reduced from 2.6$\sigma$ to 1.7$\sigma$). The inclusion of the $H_0$ measurement in the combination of data sets, together with the consideration of $N_{\rm eff}$ in the cosmological model, leads to a tremendous right shift of $n_s$ (see the comparison of the green and blue contours in Fig.~\ref{fig6}), which greatly changes the situation of the inflation model selection. Future experiments on accurately measuring the Hubble constant and searching for light relics (dark radiation) would further test the robustness of our result in this paper.

\section{Conclusion}\label{sec:5}

In this paper, we investigate how the constraints on the inflation models are affected by considering the latest local measurement of the Hubble constant in the cosmological global fit. We constrain the primordial power spectra of both scalar and tensor perturbations by using the current cosmological observations including the Planck 2015 CMB full data, the BICEP2 and Keck Array CMB B-mode data, the BAO data, and the direct measurement of $H_0$. In order to relieve the tension between the local determination of the Hubble constant and the other astrophysical observations, we consider the additional parameter $N_{\rm eff}$ in the cosmological model. We make comparison for the $\Lambda$CDM+$r$, $\Lambda$CDM+$r$+$N_{\rm eff}$, $\Lambda$CDM+$r$+$d n_{\rm s}/d \ln k$, and $\Lambda$CDM+$r$+$d n_{\rm s}/d \ln k$+$N_{\rm eff}$ models.

We find that the inclusion of $N_{\rm eff}$ indeed effectively relieves the tension. Comparing the $\Lambda$CDM+$r$ and $\Lambda$CDM+$r$+$N_{\rm eff}$ models, the tension is reduced from 2.6$\sigma$ to 1.7$\sigma$. The comparison also shows that the addition of one parameter, $N_{\rm eff}$, leads to the decrease of $\chi^2$ by 4.804. When the running of the spectral index $d n_{\rm s}/d \ln k$ is considered, we find that the fit results are basically not changed and $d n_{\rm s}/d \ln k=0$ is well consistent with the current data. Therefore, it is meaningful to consider the $\Lambda$CDM+$r$+$N_{\rm eff}$ model when the latest measurement of the Hubble constant is included in the analysis.

We constrain the $\Lambda$CDM+$r$+$N_{\rm eff}$ model using the current Planck+BK+BAO+$H_{0}$ data. We find that, in this case, the scale invariance is only excluded at the 3.3$\sigma$ level and $\Delta N_{\rm eff}>0$ is favored at the 1.6$\sigma$ level. We then compare the obtained 1$\sigma$ and 2$\sigma$ contours of $(n_s,r)$ with the theoretical predictions of some selected typical inflation models. We find that, in this case, both the convex and concave potentials are favored at the 2$\sigma$ level, although the concave potential is more favored. The natural inflation model is now excluded at more than 2$\sigma$ level, the Starobinsky $R^2$ inflation model becomes only favored at around 2$\sigma$ level, and the most favored model becomes the SBS inflation model.

%Note that our result is only a reference, not the final conclusion. If more accurate observational information is obtained from the CMB measurement, the direct $H_{0}$ measurement and other cosmological observations in future, it will be beneficial for pinning down these constraints on inflation models.

\begin{acknowledgments}

This work was supported by the National Natural Science Foundation of China (Grants No.~11522540 and No.~11690021), the National Program for Support of Top-notch Young Professionals, and the Provincial Department of Education of Liaoning (Grant No.~ L2012087).

\end{acknowledgments}

\end{document}